\documentclass[5p]{elsarticle}
\usepackage{graphicx}
\usepackage{dcolumn}
\usepackage{bm}
\usepackage{color}
\usepackage{ascmac}
\usepackage{amssymb}
\usepackage{amsmath}
\usepackage[dvipdfmx,hypertex]{hyperref}

\makeatletter
\def\ps@pprintTitle{%
 \let\@oddhead\@empty
 \let\@evenhead\@empty
 \def\@oddfoot{}%
 \let\@evenfoot\@oddfoot}
\makeatother

\begin{document}


\title{Five dimensional $O(N)$-symmetric CFTs from conformal bootstrap}


\author{Yu Nakayama}
\author{Tomoki Ohtsuki}
\address{Kavli Institute for the Physics and Mathematics of the Universe (WPI), \\ Todai Institutes for Advanced Study,
University of Tokyo, \\ 
5-1-5 Kashiwanoha, Kashiwa, Chiba 277-8583, Japan} 

\date{\today}

\begin{abstract}
We investigate the conformal bootstrap approach to $O(N)$ symmetric CFTs in five dimension with particular emphasis on the lower bound on the current central charge. The bound has a local minimum for all $N>1$, and in the large $N$ limit we propose that the minimum is saturated by the critical $O(N)$ vector model at the UV fixed point, the existence of which has been recently argued by Fei, Giombi, and Klebanov. The location of the minimum is generically different from the minimum of the lower bound of the energy-momentum tensor central charge when it exists for smaller $N$.
 To better understand the situation, we examine the lower bounds of the current central charge of $O(N)$ symmetric CFTs in three dimension to compare. We find the similar agreement in the large $N$ limit but the discrepancy for smaller $N$ with the other sectors of the conformal bootstrap.

\end{abstract}

\maketitle 
\section{Introduction} 
The recent technological advancement in the conformal bootstrap program has made it an indispensable tool in the non-perturbative study of higher-dimensional ($d > 2$) conformal field theories (CFTs). The central idea is to examine the constraints on correlation functions from their basic requirements such as unitarity and crossing symmetry \cite{Ferrara:1973yt}\cite{Polyakov:1973ha}, which has been put into practical use recently \cite{Rattazzi:2008pe}. By truncating the problem to a finite convex-optimization one, they succeeded in deriving a rigorous bound on the conformal dimensions of certain operators without assuming the explicit form of the Hamiltonian (or Lagrangian). Soon after, the technique has been generalized to constrain other parameters in CFTs such as operator product expansion coefficients \cite{Caracciolo:2009bx} and central charges \cite{Poland:2010wg}\cite{Rattazzi:2010gj}. 

While these constraints are non-trivial and their complete generality is a significant achievement, what is astonishing is that a certain class of interacting CFTs, whose existence is known from the other methods, seem to show up as a singular behavior of the bound such as ``kinks'' saturating the constraint there \cite{Rychkov:2009ij}\cite{Poland:2011ey}\cite{El-Showk:2012ht}\cite{Beem:2013qxa}\cite{Kos:2013tga}\cite{El-Showk:2013nia}\cite{El-Showk:2014dwa}\cite{Nakayama:2014lva}. Although the fundamental reason for such phenomena is not yet clear to us, we are convinced that these CFTs do lie at the
extreme corners of the conformal bootstrap constraints. In particular, the hypothesis that there exists a unitary CFT at the singular point of the bound has led to the most numerically accurate prediction for the critical exponents of the $3d$ Ising model in \cite{El-Showk:2014dwa}.

Meanwhile the previous studies have been focusing on space-time dimensions $d\le 4$. 
One underlying theoretical presumption 
may be that we have much less knowledge for CFTs in $d > 4$, where in the Hamiltonian (or Lagrangian) formulation,
it is harder to prescribe stable (i.e. bounded from below), UV-complete, and unitary interactions. Despite the difficulty, in a recent paper \cite{Fei:2014yja} they have argued that there exist $O(N)$-symmetric CFTs in $ 4<d<6$ dimension based on the $\epsilon$ expansion in $d=6-\epsilon$ dimension, which can be thought of as analogues of the critical $O(N)$ vector models in $2<d <4$ dimension.

These models, if they exist in particular in $d=5$ dimension, are of physical interest in several contexts. The first is their unusual asymptotic safe behavior under the renormalization group flow, where these theories can be characterized as UV interacting fixed points of the $O(N)$ symmetric Hamiltonian, rather than the more conventional IR ones. The second is the AdS$_{6}$/CFT$_{5}$ correspondence, where the dual theories in the AdS$_6$ bulk include higher-spin fields in the large $N$ limit as the dual theories of critical $O(N)$ vector models in $d=3$ dimension did in the AdS$_4$/CFT$_3$ correspondence \cite{Klebanov:2002ja} (for a recent review, see e.g. \cite{Giombi:2012ms}). 
However, setting $\epsilon =1$ in the asymptotic $\epsilon$ expansion would require careful treatment and na\"ive expectation might fail.
It is highly desirable to provide further evidence from non-perturbative methods such as the conformal bootstrap.

In this paper, we perform the conformal bootstrap program for $O(N)$ symmetric CFTs in $d=5$ dimension. 
The large $N$ prediction tells us that unlike what happens in $d=3$ dimension studied in \cite{Kos:2013tga}, the proposed fixed point of the $O(N)$ vector models in $d=5$ dimension cannot saturate the bound for the conformal dimensions of the first scalar operator in the singlet as well as the symmetric traceless tensor representation. Our strategy therefore is to investigate the other quantities such as the energy-momentum tensor central charge and the current central charge instead.

The organization of the paper is as follows. In section 2, we collect some relevant aspects of the conformal fixed points proposed in \cite{Fei:2014yja} and the conformal bootstrap program with the $O(N)$ symmetry. In section 3, we present our numerical results of the conformal bootstrap in $d=5$ dimension. In section 4 we present the results for the analogous computation in $d=3$ dimension for comparison. In section 5, we interpret our results with further discussions.

\section{Large $N$ prediction and conformal bootstrap}
In the Hamiltonian formulation, the critical $O(N)$ vector model is realized as the universality class of the $O(N)$ symmetric Hamiltonian:
\begin{align}
\mathcal{H} = (\partial_\mu \phi_i)(\partial_\mu \phi_i) + \lambda (\phi_i \phi_i)^2 ,  \label{phi4}
\end{align} 
where $i = 1,\cdots N$. This Hamiltonian admits a natural $\epsilon$ expansion in $ d= 4-\epsilon$ dimension with the perturbative IR fixed point for $\epsilon > 0$, which is identified with the Heisenberg fixed point in $d=3$ dimension (at $\epsilon = 1$). Our focus is the unconventional regime: $\epsilon < 0$ ($\epsilon = -1$ or $d=5$ eventually). The Hamiltonian becomes non-renormalizable in the power-counting sense, yet we may expect the existence of an asymptotic safe UV fixed point. The perturbative UV fixed point for small negative $\epsilon$, however, leads to a negative value of $\lambda$, which may call for the question of instability while the anomalous dimensions still appear to be consistent with the unitarity bound.

On the other hand, we may perform the large $N$ expansion of \eqref{phi4}. The analysis made perfect sense in $4<d<6$ as well as in $2<d<4$. For sufficiently large $N$, the known properties of the hypothetical UV fixed point do not contradict with the unitarity constraints of the CFT.

For later purposes, we collect the large $N$ predictions for critical $O(N)$ vector models in $d=3$ and $d=5$ dimensions available in the literature \cite{Vasiliev:1982dc}\cite{2002PhRvE..66b7102G} (see also \cite{Kos:2013tga}\cite{Fei:2014yja}). The conformal dimension of the scalar operator in $O(N)$ vector representation (i.e. $\phi_i$ in \eqref{phi4}) is
\begin{align}
\Delta_{\phi}^{d=3} &= \frac{1}{2} + \frac{0.135095}{N} - \frac{0.0973367}{N^2} -\frac{0.940617}{N^3} + \cdots \cr
\Delta_{\phi}^{d=5} &= \frac{3}{2} + \frac{0.216152}{N} - \frac{4.342}{N^2} - \frac{121.673}{N^3} + \cdots.\label{eq:largeN-Deltaphi}
\end{align}
For the first scalar operator in the $O(N)$ singlet (S) representation,
\begin{align}
\Delta_{S}^{d=3} &= 2 -\frac{1.08076}{N} - \frac{3.0476}{N^2} + \cdots \cr
\Delta_{S}^{d=5} &= 2 + \frac{10.3753}{N} + \frac{206.542}{N^2} + \cdots .
\end{align}
For the $O(N)$ symmetric traceless tensor (T) representation, 
\begin{align}
\Delta_{T}^{d=3} &= 2\Delta_{\phi} +\frac{0.810569}{N} + \frac{0}{N^2} + \cdots \cr
\Delta_{T}^{d=5} &= 2\Delta_{\phi} -\frac{2.16152}{N} + \frac{16.5083}{N^2} + \cdots.
\end{align}
As observed in \cite{Fei:2014yja}, the asymptotic behavior of the $1/N$ expansion in $d=5$ dimension is significantly worse than in $d=3$ dimension. Assuming the $O(1/N^3)$ prediction to be compatible with the unitarity bound $\Delta_{\phi} \ge 3/2$ , \cite{Fei:2014yja} predicted that unitary conformal fixed points exist for $N > 35$ in $d=5$ dimension.

$1/N$ corrections to the energy-momentum tensor central charge and the current central charge are available in \cite{Petkou:1995vu}:
\begin{align}
\frac{C_T^{d=3}}{C_{T}^{\mathrm{free},d=3}} &= 1 -\frac{0.450316}{N} + \cdots \cr
\frac{C_T^{d=5}}{C_{T}^{\mathrm{free},d=5}} &= 1 -\frac{0.0905669}{N} + \cdots
\label{eq:largeN-EMCC}
\end{align}
and
\begin{align}
\frac{C_J^{d=3}}{C_{J}^{\mathrm{free},d=3}} &= 1 -\frac{0.720506}{N} + \cdots \cr
\frac{C_J^{d=5}}{C_{J}^{\mathrm{free},d=5}} &= 1 -\frac{0.461124}{N} + \cdots,\label{eq:largeN-CurrentCC}
\end{align}
which we have normalized by the values at the Gaussian fixed point.

In \cite{Fei:2014yja}, they have proposed an alternative UV complete description based on the Hamiltonian
\begin{align}
\mathcal{H} &= (\partial_\mu \phi_i)(\partial_\mu \phi_i) + (\partial_\mu \sigma)(\partial_\mu \sigma) +g_1 \sigma(\phi_i \phi_i) + g_2 \sigma^3 \ , \label{six}
\end{align}
where $\sigma$ is an $O(N)$ singlet. The Hamiltonian \eqref{six} allows a perturbative search of the fixed point in $d = 6-\epsilon$ dimension. For sufficiently small $\epsilon$, they found that for large $N \ge 1039$, there is an IR stable fixed point corresponding to the critical $O(N)$ vector models we have described above. They also found the other (unstable) fixed points for any $N$ that do not have the conventional $1/N$ expansions (rather $1/\sqrt{N}$ expansions).

In the following, we would like to investigate the conformal bootstrap program for $O(N)$ symmetric CFTs in $d=5$ dimension. Before doing any numerical analysis, we immediately realize that in the large $N$ limit the critical $O(N)$ vector models in $d=5$ dimension cannot be seen as a ``kink" in the bound for the conformal dimensions in either S or T sector (unlike the study in $d=3$ dimension \cite{Kos:2013tga}) because the conformal dimensions of these operators in the large $N$ limit are predicted to be smaller than those in the generalized free theories. Indeed, the preliminary numerical conformal bootstrap confirms this.
Our strategy, therefore, is to study the lower bound of the energy-momentum tensor central charge and the current central charge.

Before presenting the results, let us briefly comment on the conformal bootstrap program for CFTs with global $O(N)$ symmetry, assuming the presence of a scalar field $\phi _i $ in the $O(N)$ vector representation. The detailed prescription to obtain the lower bound of the energy momentum tensor central charge from the four point function $\langle \phi _{i_1} (x_1) \phi _{i_2} (x_2) \phi _{i_3} (x_3) \phi _{i_4}(x_4) \rangle$ has been given in \cite{Kos:2013tga} for any $d$, so we will not delineate them here, and we follow their notation. See also \cite{Rattazzi:2010yc}\cite{Vichi:2011ux}\cite{Poland:2011ey} for further details on the bootstrap program with general global symmetries. To obtain the lower bound of the current central charge, all we have to do is to change the normalization point of the linear functional $\Lambda$ from the energy-momentum tensor to the $O(N)$ current \cite{Poland:2011ey} (see \cite{Berkooz:2014yda} for a similar approach to current central charges in $4d$ $\ \mathcal{N}=1$ supersymmetric theories). In the actual computation, the intermediate states with spin $l \le 20$ and $l=1001 ,\ 1002$ are taken into account (the latter is to include the asymptotic effects), and no assumption is made for the spectra other than the unitarity bound unless otherwise stated. To obtain a partial fractional approximation of the conformal blocks, we employed the algorithm of \cite{Hogervorst:2013kva} to generate the residues. All the results in this paper are derived from $k=10$ or $55\times 3$-dimensional search space which was sufficient in \cite{Kos:2013tga} to see the kinks corresponding to the critical $O(N)$ vector models in $d=3$ dimension. Our \verb+sdpa-gmp+\cite{sdpa1}\cite{sdpa2} implementation is the same as in \cite{Kos:2013tga}, except that the parameter \verb+precision+ is 350 here.

\section{Central charge bounds in $d=5$}
\begin{center}
\begin{figure}[h!!]
  \centering
  \includegraphics[width=8cm]{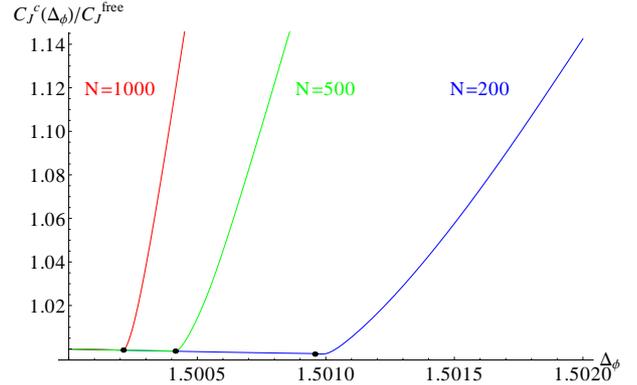}
  \caption{The lower bounds of the current central charge $C_J ^c (\Delta _\phi)$ for $d=5$ $O(N)$ symmetric CFTs with $N= 200, 500,1000$. The dots are the large $N$ predictions of the $(\Delta _\phi ,C_J )$ in (\ref{eq:largeN-Deltaphi}) and (\ref{eq:largeN-CurrentCC}).}
  \label{fig:1}
\end{figure}
\end{center}

Let us now present our numerical results of the conformal bootstrap for the lower bound of the current central charge with $N= 200, 500, 1000$ in Figure \ref{fig:1} along with the prediction of the $\Delta _\phi $ and $C_J$ from the large $N$ expansion in (\ref{eq:largeN-Deltaphi}) and (\ref{eq:largeN-CurrentCC}). As we can see, the curves show sudden changes in their slopes after their local minima, and for $N=1000$ and $N=500$, their locations agree with the large $N$ predictions within our horizontal precisions, $2\times 10^{-5}$. For $N=200$ the behavior is similar, but there seems to be a small discrepancy between the location of the minimum and the large $N$ prediction. At this point, it is not clear to us if this discrepancy comes from the poor asymptotic $1/N$ expansions in $d=5$ dimension, or the possibility that the minimum does not correspond to the actual CFT.

\begin{center}
\begin{figure}[h!!]
  \centering
  \includegraphics[width=8cm]{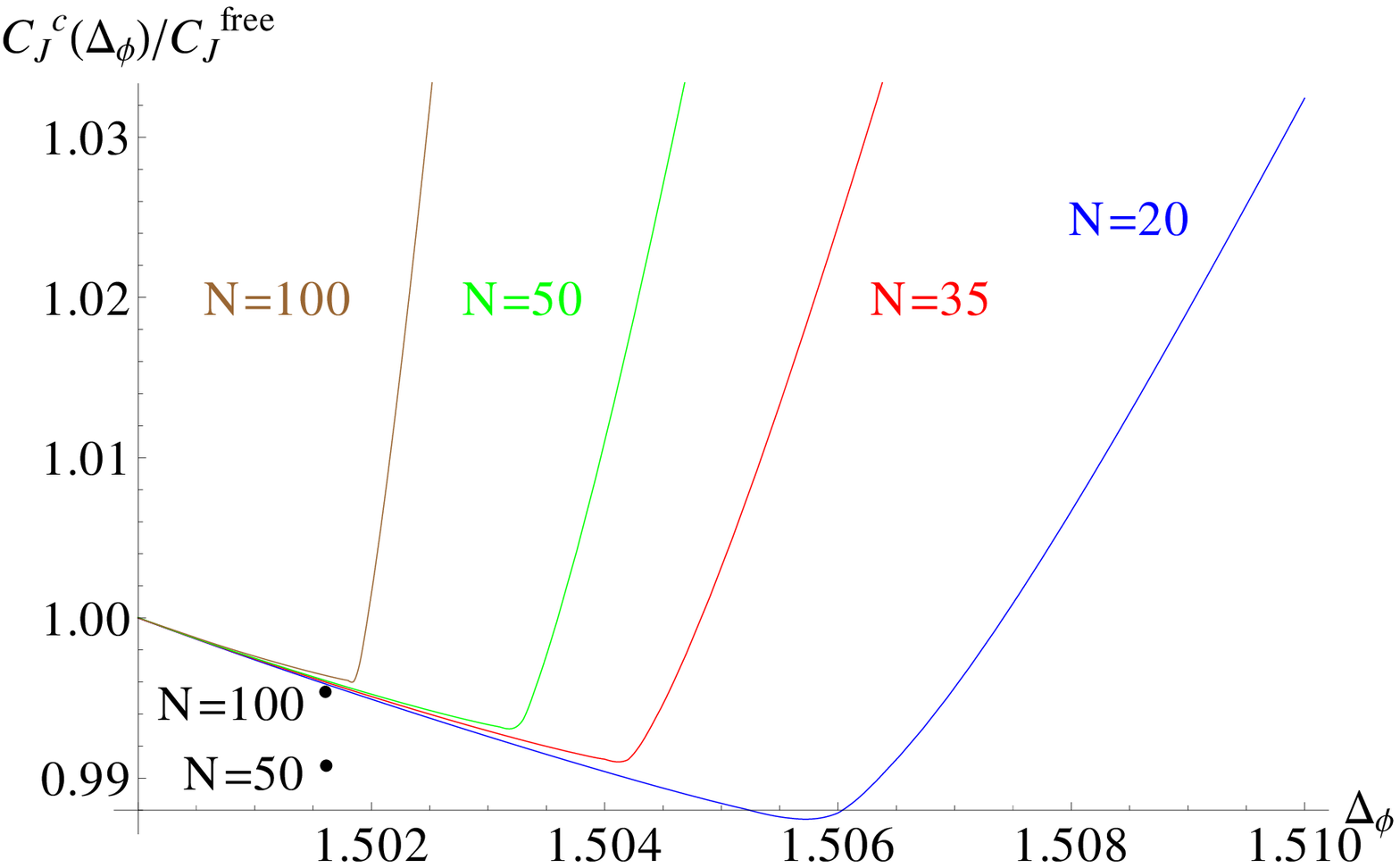}
  \caption{The lower bounds of the current central charge $C_J ^c (\Delta _\phi)/C_J ^{\mathrm{free}}$ for $d=5$ $O(N)$ symmetric CFTs with $N= 20, 35,50, 100$. The dots are the large $N$ predictions of the $\Delta _\phi$.}
  \label{fig:2}
\end{figure}
\end{center}

\begin{center}
\begin{figure}[h!!]
  \centering
  \includegraphics[width=8cm]{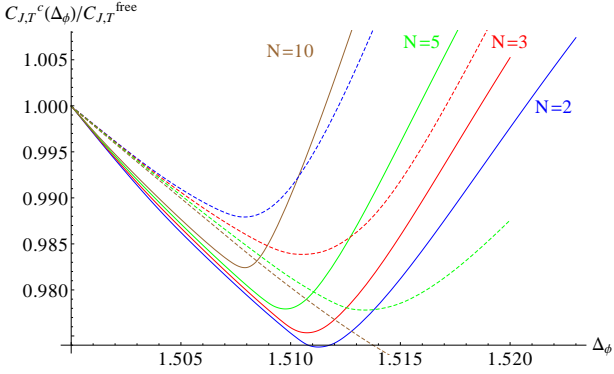}
  \caption{$C_{J,T} ^c (\Delta _\phi)/C_{J,T} ^{\mathrm{free}}$ for $d=5$ $O(N)$ symmetric CFTs with $N= 2, 3,5, 10$. The solid lines are the lower bounds of the current central charge while the dashed lines are those of the energy-momentum tensor central charge with the corresponding colors. }
  \label{fig:3}
\end{figure}
\end{center}

For reference we note that substituting $N=200$ into the $1/N^3$ term in (\ref{eq:largeN-Deltaphi}) gives the value $-1.5\times 10^{-5}$, which is comparable with the discrepancy here.
We continue to present the results for smaller values of $N$ in Figure \ref{fig:2}. The disagreement between the locations of the local minima and the large $N$ predictions becomes even worse. While in \cite{Fei:2014yja} they claim the absence of the unitary conformal fixed points for $N\le 35$, we do not find any qualitative changes of the curves. In particular the local minima persist for any $N$.
Finally, we present the lower bounds of the current central charge for $N = 2,3,5,10$ in Figure \ref{fig:3} along with those of the energy-momentum tensor central charge.
For these values of $N$, we also find non-trivial local minima for the lower bounds of the energy-momentum tensor central charge at the range close to where the lower bounds of the current central charge show the local minima. As we can see, the locations of the minima for the lower bounds of the energy-momentum tensor central charge do not necessarily coincide with those of the current central charge.

The large $N$ asymptotic slope for the bounds of the current central charge in the smaller $\Delta_{\phi}$ region agrees with the large $N$ predictions from \eqref{eq:largeN-Deltaphi} and \eqref{eq:largeN-CurrentCC}. However, the similar asymptotic slope for the lower bounds of the energy-momentum tensor central charge disagrees with the large $N$ predictions of \eqref{eq:largeN-Deltaphi} and \eqref{eq:largeN-EMCC}. We will give possible interpretations of these facts in section 5.

One may further attempt to improve the above results by assuming additional conditions for the intermediate states in the conformal bootstrap. 
For example, in the literature \cite{El-Showk:2012ht}\cite{Kos:2013tga}\cite{El-Showk:2014dwa}, the conditions $\Delta_{S,T} \ge 1$ for spin $0$ intermediate states  were imposed in $d=3$ dimension while the unitarity bound is weaker: $\Delta_{S,T} \ge 1/2$. The additional assumption should be consistent with the CFTs we  would like to find:  we know that the available spectra for the critical $O(N)$ vector models (either from the other sectors of the conformal bootstrap or from the other methods)
satisfy $\Delta_{S,T} \ge 1$, so it should make the bound stronger without excluding them. 

\begin{center}
\begin{figure}[h!!]
  \centering
  \includegraphics[width=8cm]{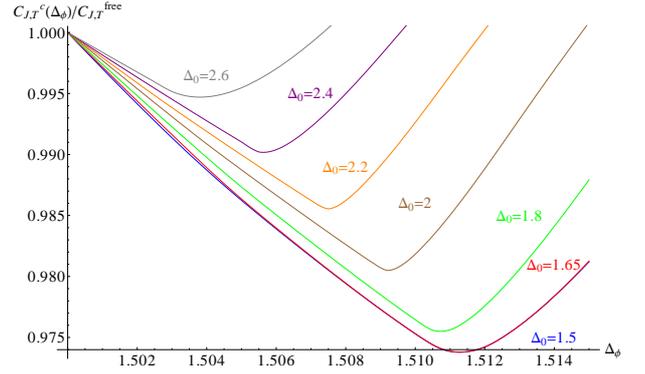}
  \caption{$C_J ^c (\Delta _\phi)/C_J ^{\mathrm{free}}$ for $d=5$ $O(2)$ symmetric CFTs obtained by assuming $\Delta_{S,T} \ge \Delta_0$ with $\Delta _0$ running over $1.5, 1.65, 1.8, 2, 2.2, 2.4, 2.6$.}
\label{fig;5dSO2-VaryingCutoff}
\end{figure}
\end{center}

In our conformal bootstrap approach to the critical $O(N)$ vector models in $d=5$ dimension for smaller $N$, however, we have less knowledge of the operator contents of the CFTs we are looking for, nor there seems no other bootstrap sectors that give the prediction of the spectra (as far as we have tried). Therefore, a priori, we do not know what kind of extra assumptions make the bound stronger without excluding the non-trivial CFTs.
For an experiment, we have derived the lower bounds of the current central charge in the $O(2)$ symmetric CFTs with the assumptions $\Delta_{S,T} \ge \Delta _0$ for spin $0$ intermediate states by changing $\Delta_0$, whose results are shown in Figure \ref{fig;5dSO2-VaryingCutoff}. The bound is rather stable against shifting $\Delta _0$ from $1.5$ to $1.65$ and then starts to move. However such a behavior does not immediately imply that there is an actual CFT at the observed  minimum saturating the lower bounds for $\Delta_0 = 1.5$ or $1.65$ with a spin $0$ intermediate state whose conformal dimension is $\Delta_{S,T} < 1.8$ . To see what is happening, we will compare the situations in $d=3$ dimension in the next section.

\section{Current central charge bounds in $d=3$}
In order to better understand the situations in $d=5$ dimension, we have performed the similar analysis of the lower bounds of the current central charge in $d=3$ dimension for $O(N)$ symmetric CFTs. In particular, we would like to address the question if the local minima of the lower bounds of the current central charge can be associated with the critical $O(N)$ vector models.

\begin{center}
\begin{figure}[h!!]
  \centering
  \includegraphics[width=8cm]{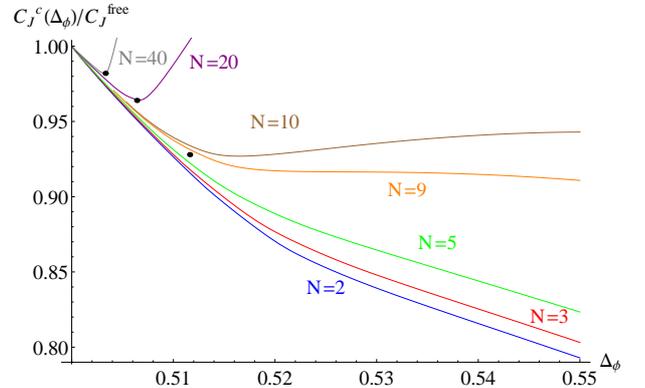}
  \caption{$C_J ^c (\Delta _\phi)/C_J ^{\mathrm{free}}$ for $d=3$ $O(N)$ symmetric CFTs with $N= 2, 3,5,9,10,20,40$. The bounds are completely general -- i.e. no assumption other than the unitarity bound is made. The dots are the large $N$ predictions of $(\Delta_\phi , C_J)$ for $N=40, 20,10$ critical vector models from the left.
  \label{fig:4}}
\end{figure}
\end{center}

To keep the story in parallel with that in $d=5$ dimension, we have first derived the lower bounds from the conformal bootstrap program without assuming any additional conditions for the spectra of the intermediate states other than the unitarity bound (which is $1/2$ for spin $0$ operators in $d=3$ dimension). For sufficiently large $N$, Figure \ref{fig:4} shows that the lower bounds of the current central charge possess the local minima as in $d=5$ dimension, and their locations in the large $N$ limit coincide with the large $N$ predictions of $\Delta_{\phi}$ of the critical $O(N)$ vector models \eqref{eq:largeN-Deltaphi}.
However, for smaller $N$, we see that the location of the minimum begins to deviate from the $\Delta_{\phi}$ predicted in the other sectors of the conformal bootstrap (e.g. S or T sector) in \cite{Kos:2013tga}. Furthermore, for $N < 9$, the minimum of the lower bound of the current central charge disappears.

\begin{center}
\begin{figure}[h!!]
  \centering
  \includegraphics[width=8cm]{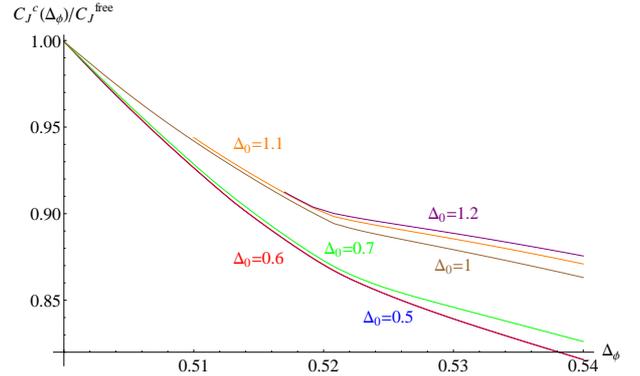}
  \caption{$C_J ^c (\Delta _\phi)/C_J ^{\mathrm{free}}$ for $d=3$ $O(2)$ symmetric CFTs with assuming $\Delta_{S,T} \ge \Delta_{0}$ for the spin $0$ intermediate states, with $\Delta_0$ running over $0.5, 0.6, 0.7,1,1.1,1.2$.}
\label{fig:3dSO2-VaryingCutoff}
\end{figure}
\end{center}

For comparison we note that the location of the local minimum of the lower bound of the energy-momentum tensor central charge for $N>1$, if any, does not coincide with either the $\Delta_{\phi}$ predicted from the kinks in the $S$ and $T$ sectors or the minimum of the lower bound of the current central charge we obtained (see FIG 4 of \cite{Kos:2013tga})\footnotemark. In \cite{El-Showk:2014dwa}, the energy-momentum tensor central charge was successfully chosen as a good trial function for the precision conformal bootstrap program in the $3d$ Ising model. Our results suggest that neither the energy-momentum tensor central charge nor the current central charge by themselves may be good candidates for the precision conformal bootstrap of the critical $O(N)$ vector models in $d=3$ dimension except in the large $N$ limit for the latter.
\footnotetext{One can improve the situation by using further information available from the other sectors in the conformal bootstrap program. For example, in deriving  FIG 5 of \cite{Kos:2013tga}, they assumed $\Delta_S \ge \Delta _S ^c (\Delta _\phi)$ where $\Delta _S ^c$ is the {\it upper} bounds on the first operator in the $S$ sector (i.e. the curve shown in FIG 2 of \cite{Kos:2013tga}). However, in $d=5$ dimension we cannot import such information at present because the critical $O(N)$ vector models are sitting well-below the bounds for the $S$ sector.}
As stated at the end of the previous section, one may attempt to improve these bounds by examining the conformal bootstrap program with additional conditions, i.e. $\Delta_{S,T} \ge \Delta_0$ rather than the unitarity bound $\Delta_{S,T} \ge 1/2$. 
While we have some knowledge of the spectra of operators here in $d=3$ dimension, we again vary $\Delta_0$ to compare the situations in $d=5$ dimension. The results for the lower bound of the current central charge in the $O(2)$ symmetric CFTs are presented in Figure \ref{fig:3dSO2-VaryingCutoff}. The behavior under changing $\Delta_0$ is similar to that in $d=5$: at first stage it is rather stable and then starts to move, and it begins to show a ``kink".

 Although the bounds with the assumption $\Delta _0 =1$ and $1.1$ show a rather sharp ``kink'' around the values which we have expected in the critical $O(2)$ vector model from our a-priori knowledge, we are forced to conclude that the critical $O(2)$ vector model saturates {\it none} of these bounds because the other methods \cite{Campostrini:2006ms}\cite{Kos:2013tga}  strongly  suggest that
the critical $O(2)$ vector model does satisfy $\Delta_{S,T} \ge 1.2$ and thereby its current central charge must be bounded from below by the curve with $\Delta_0 = 1.2$. Our results then imply that without a further input from the other sectors to tune the assumptions of the intermediate states, 
it seems difficult to use the current central charge itself as a trial function for the precision conformal bootstrap of the critical $O(N)$ vector models.

\section{Discussions}
In this paper, we have studied the conformal bootstrap program for $O(N)$ symmetric CFTs in $d=5$ dimension with particular emphasis on the lower bound on the current central charge. In the large $N$ limit, given the agreement with the large $N$ analysis together with the more solid results in $d=3$ dimension, we are confident that the observed local minima correspond to the critical $O(N)$ vector models discussed in \cite{Fei:2014yja}.

We, however, did not find any evidence for the conformal window proposed in \cite{Fei:2014yja}. There are a couple of possible scenarios here.
(1) The conformal window does not exist: the critical $O(N)$ vector models are unitary CFTs for any $N>1$. (2) At finite $N$, there exist crossovers with the other fixed points such as the ones predicted in \cite{Fei:2014yja} from the $\epsilon$ expansion in $d = 6-\epsilon$ dimension. 
(3) The conformal window does exist, but the lower bound of the current central charge did not capture it: the observed local minima do not correspond to any unitary CFT.

In the first scenario, the discrepancy for finite $N$ in $d=5$ dimension could be explained by the poor asymptotic behavior of the large $N$ expansion.
In the second scenario, the discrepancy between the location of the local minimum in the lower bound of the energy-momentum tensor central charge and that of the current central charge could be understood as the existence of two (or more) distinct conformal fixed points within the same universality class. It would also explain the observed asymptotic slope of the lower bound of the energy-momentum tensor central charge that differs from the large $N$ prediction.
In the third scenario, the possible absence of CFTs at the local minima of the bounds may have the same origin of the situations in $d=3$ dimension for smaller $N$ where the critical $O(N)$ vector models do not saturate the bounds of the current central charge at the minima (without tuning the extra assumptions for the intermediate states). 

It would be interesting to see if these puzzles in smaller $N$ can be resolved within the conformal bootstrap program. Alternatively, to test these scenarios directly in $d=5$ dimension independently from the conformal bootstrap, we may perform the higher loop $\epsilon$ expansions with a careful resummation.

Finally, since our results appear to be robust in the large $N$ limit, 
the analysis of higher-spin current operators from the conformal bootstrap may be of practical use to establish its connection to the AdS$_6$/CFT$_5$ correspondence with higher-spin fields propagating in the bulk. 

\section*{Acknowledgements}
This work is supported by the World Premier International Research Center Initiative (WPI Initiative), MEXT. T.O. is supported by JSPS Research Fellowships for Young Scientists and the Program for Leading Graduate Schools, MEXT.

\providecommand{\href}[2]{#2}\begingroup\raggedright
\endgroup
\end{document}